# Some peculiarities of water freezing at small sub-zero temperatures

Alexei V. Finkelstein[1,2]

[1]Institute of Protein Research of the Russian Academy of Sciences, 142290 Pushchino, Moscow Region, Russia.
[2]Biotechnology Department of the Lomonosov Moscow State University, 4 Institutskaya Str., 142290 Pushchino, Moscow Region, Russia.

E-mail: afinkel@vega.protres.ru

**I consider the kinetics of water freezing and show that, at small sub-zero temperatures, (i) the time of ice nucleation within the bulk water environment is enormous and therefore cannot take place either in lakes of in living cells; (ii) that the ice nucleation needs some ice-binding surfaces to occur, but (iii) even this kind of ice nucleation can take place, as a rule, only at the temperatures that are a few degrees below 0°C. Further, I discuss factors that can drastically reduce the ice nucleation time at nearly-zero temperatures both in open reservoirs, where water contacts with air, and in cells, where there is no such contact.**

**Introduction**

As known, water freezes at 0°C. More precisely, 0°C is the temperature at which ice coexists with water. However, it is also known that the ice-water transition is a first-order phase transition, so that water at a sub-zero temperature can exist, sometimes for an extremely long time, in the supercooled state, since the nucleation of a solid phase in a liquid (e.g., ice in pure water) can be very slow [1, 2].

In this work, I first in general evaluate the characteristic time of nucleation of the "new", crystalline phase (i) within and (ii) on the surface of the "old" liquid one near the point of thermodynamic equilibrium of these two phases, and then discuss factors that can drastically reduce this, often a huge time in different systems.

The formation of an ordered phase from a melt is considered in books [1-4]. For a three-dimensional (3D) crystal this is a first-order phase transition. The emergence of an ordered 2D layer is not, strictly speaking, the first-order phase transition, but this ordered layer is quite similar to a crystal (and another 2D phase to a melt) [5], which allows estimating the free energy of intermediates of its formation in the same way as this is done for nucleation and growth of a 3D crystal. The one-dimensional system can also include quite distinct ordered and disordered phases, and, though the emergence of an ordered 1D phase is not the first-order phase transition [6], it allows estimating the free energy of intermediates of formation of the ordered 1D phase.

The free energy change during the formation of a piece of the compact $d$-dimensional ($d = 3$, 2 or 1) new phase (see Figs. 1, 2) consisting of $n$ particles ($n \geq 2$) can be approximately estimated as

$$G(n) \approx n\Delta\mu + \alpha_d n^{\frac{d-1}{d}} B_d, \qquad (1)$$

where $\Delta\mu < 0$ is the chemical potential of a molecule in the "new", arising phase minus that in the "old" one (so that at the point of thermodynamic equilibrium of phases $\Delta\mu = 0$), $B_d > 0$ is the additional free energy of one molecule on the surface (for $d=3$), perimeter (for $d=2$) or end (for $d=1$) of the "new" phase, and $\alpha_d n^{\frac{d-1}{d}}$ is the number of molecules on the surface (for $d=3$) or perimeter (for $d=2$) or end (for $d=1$) of a compact piece of the new phase consisting of $n$ particles (at $d = 1$: $\alpha_1 = 2$; at $d = 2$: $\alpha_2 = \sqrt{4\pi} = 3.54$ for a circle, and $\alpha_2 = 2d = 4$ for a square; at $d = 3$: $\alpha_3 = \sqrt[3]{36\pi} \approx 4.84$ for a sphere, and $\alpha_3 = 2d = 6$ for a cube). The free energy of non-compact (with large $\alpha_d$) intermediates of new 2D and 3D phases is higher than that of the compact ones; thus, when estimating the time of nucleation of a new, crystalline phase, we may ignore slow paths going through the non-compact structures.

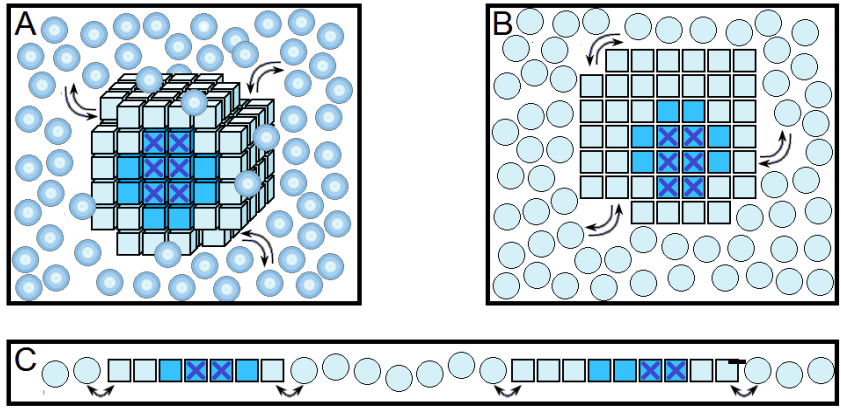

FIG. 1. Schematic presentation of a slice of a compact structured piece of the three-dimensional phase (A), of a compact structured piece of the two-dimensional phase (B), and of a mixture of structured and disordered pieces of the one-dimensional phase. The pieces of the "new", structured phases (their particles are shown in cubes on panel A and squares on panels B, C) were formed from the "free" particles of a liquid (shown in balls on panel A and in circles on panels B, C). Dark blue cubes and squares mark the "seeds" of the structured pieces, that is, the smallest minimally stable (as compared to liquid) parts of the structured pieces; light blue cubes and squares are particles that stuck to these "seeds" later. Dark blue crosses mark the "nuclei" of the "seeds", that is, their pieces that had minimal stability along the pathway of the structuring process.

**Nucleation**

The nucleation of a new phase is a multi-step (Fig. 2) reaction in multi-dimensional systems which are of the main interest for us.

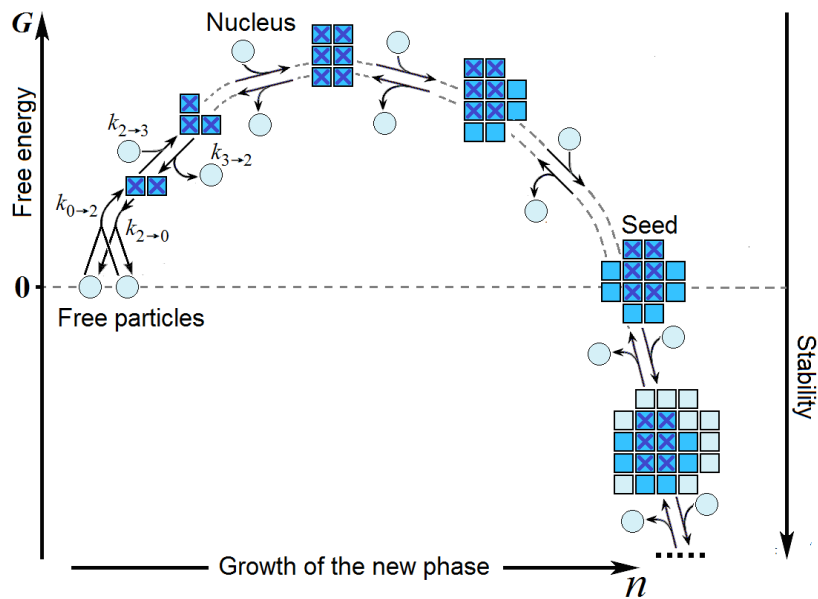

FIG. 2. Schematic presentation of formation of a crystal from free particles ina multi-dimensional system. It begins with sticking together of a few particles in a configuration that allows further growth of the ordered phase. However, the rate of appearance of the new phase is determined not by this first step; it is determined by formation of the "nucleus" of this phase that corresponds to the activation barrier, i.e., has the minimal stability in the course of this process (as in Fig. 1, this piece is marked with crosses). $k_{0\to2}$; $k_{2\to3}, k_{3\to4}, ...$ are the rate constants of initiation and, then, growth of the new phase; $..., k_{4\to3}, k_{3\to2}$; $k_{2\to0}$ are the rate constants of its decrease and, then, its disappearance (see text). Other designations are the same as in Fig. 1.

It begins with sticking together of several (in Fig. 2 — two) particles in a configuration that allows further growth of this new (in Fig. 2 — two-dimensional) phase through addition of "free" particles from the melt.

We are interested in the case when the "new" phase is stable ($\Delta\mu < 0$), but the temperature is close to the phase equilibrium point ($T_0 \approx 273$ K for water/ice transition), i.e., when $-\Delta\mu \ll k_B T$ ($k_B$ being the Boltzmann constant), and $-\Delta\mu \ll B_d$.

Then, in a multi-dimensional system (i.e., at $d>1$), $G(n)$ first grows with increasing $n$, then passes through the maximum in the transition state (corresponding to the "nucleus" of the piece of a new phase), and then decreases (Fig. 2). The maximum of $G(n)$ is reached (at $d>1$; see Fig. 1A, 1B) where $\frac{dG}{dn}\big|_{n=n_0} = \Delta\mu + \frac{d-1}{d}\alpha_d n_0^{\frac{-1}{d}} B_d = 0$, i.e., in the transition state; there, the nucleus includes

$$n_0 \approx \left(\frac{d-1}{d}\alpha_d \frac{B_d}{-\Delta\mu}\right)^d \quad (2)$$

particles, and the "seed". i.e., the minimally stable (with $G(n) = 0$ at $n > 0$) piece of the "new" multi-dimensional phase contains

$$n_{\min} \approx \left(\alpha_d \frac{B}{-\mu}\right)^d = \left(\frac{d}{d-1}\right)^d n_0 \quad (2a)$$

particles.

The free energy of the transition state (at $d>1$) is

$$G^{\#}_{\text{nucl}} \equiv G(n_0) \approx \frac{\alpha_d B_d}{d}\left(\frac{d-1}{d}\alpha_d \frac{B_d}{-\Delta\mu}\right)^{d-1}. \quad (3)$$

The estimate for $d=3$ (homogeneous nucleation in the 3D space):

at $\alpha_3 = \sqrt[3]{36\pi}$: $n_0 \approx \frac{32\pi}{3}\left(\frac{B_3}{-\Delta\mu}\right)^3$, $G^{\#}_{\text{nucl}} \approx \frac{16\pi B_3}{3}\left(\frac{B_3}{-\Delta\mu}\right)^2$;

at $\alpha_3 = 6$: $n_0 \approx 64\left(\frac{B_3}{-\Delta\mu}\right)^3$, $G^{\#}_{\text{nucl}} \approx 32 B_3\left(\frac{B_3}{-\Delta\mu}\right)^2$.

The estimate for $d=2$ (heterogeneous nucleation at the 2D boundary of the 3D space):

at $\alpha_2 = \sqrt{4\pi}$: $n_0 \approx \pi\left(\frac{B_2}{-\Delta\mu}\right)^2$, $G^{\#}_{\text{nucl}} \approx \pi B_2 \left(\frac{B_2}{-\Delta\mu}\right)$;

at $\alpha_2 = 4$: $n_0 \approx 4\left(\frac{B_2}{-\Delta\mu}\right)^2$, $G^{\#}_{\text{nucl}} \approx 4 B_2 \left(\frac{B_2}{-\Delta\mu}\right)$.

So, the diameter of the "nucleus" of the new multi-dimensional phase corresponds to a row of about $(n_0)^{\frac{1}{d}} = \frac{d-1}{d}\alpha_d\left(\frac{B_d}{-\Delta\mu}\right)$ particles (cf. [2]), and the diameter of its "seed" to a row of $\approx (n_{\min})^{\frac{1}{d}} = \alpha_d\left(\frac{B_d}{-\Delta\mu}\right)$ particles.

As for the new 1D phase (Fig. 1C) formation, its nucleus contains 2 particles and has free energy $G^{\#}_{\text{nucl}} = 2(B_1 + \Delta\mu)$; and the seed of the new 1D phase consists of $2\left(\frac{B_d}{-\Delta\mu}\right)$ particles.

**The time required for the nucleation of a new phase**

Let us focus on the time required for the initiation of ice, and, for the reasons that will become clear soon, neglect the time of its growth for a while.

Then, according to the transition state theory [7, 8], the time of appearance of an ice nucleus around one given H₂O molecule can be estimated as $\sim \tau \cdot \exp\left(\frac{G^{\#}_{\text{nucl}}}{k_B T}\right)$, where $\tau$ is the time of addition of one water molecule to ice; it is no less than $\tau_0 \sim 10^{-12}$ s, the typical time of thermal vibrations at 0°C. More accurately, the value of $\tau$ is estimated as the difference between the rates of water attachment to and detachment from ice ([2], chs. 3.2, 8.2): taking into account only the main terms, one can obtain an estimate $\frac{1}{\tau} \approx \frac{1}{\tau_0}\exp\left(\frac{-\varepsilon}{2k_B T}\right)\left[\exp\left(\frac{-\Delta\mu}{k_B T}\right) - 1\right]$, where $\varepsilon$ is the energy of ice sublimation. At temperatures

$T$ close to $T_0 \approx 273°K$, $\varepsilon \approx 51$ kJ/mol [9], so that $\tau \sim 10^{-7}$ s$\cdot \left(\frac{-\Delta\mu}{k_B T_0}\right)$.

At normal (~1 atm.) pressure and $T_0 \approx 273°K$, $\Delta\mu \equiv 0$ by definition, and at the temperature $T_0 - \Delta T$ (where $\Delta T \ll T_0$), $\mu = -\Delta S_{(1)}(-\Delta T) = -\Delta H_{(1)}\left(\frac{-\Delta T}{T_0}\right)$ according to the classical equations of thermodynamics (here $\Delta S_{(1)}$ and $\Delta H_{(1)}$ are the entropy and enthalpy of water freezing per 1 molecule). Experimentally, for water, $\Delta H_{(1)} \approx -6.0$ kJ/mol $\approx -2.6 k_B T_0$ [9], so that

$$\frac{\Delta\mu}{k_B T_0} \approx \frac{-\Delta T}{100°}. \tag{4}$$

and thus

$$\tau \sim 10^{-5} \cdot \left(\frac{1°}{\Delta T}\right) \text{ s}. \tag{5}$$

*The homogeneous nucleation: ice formation in the bulk water*

To begin with, we shall consider the "homogeneous" (i.e., occurring in bulk water) ice nucleation (and show that ice cannot arise in this way until the temperature is above -40°C), and then turn to the "heterogeneous" (i.e., occurring at a water boundary) ice nucleation.

An experimental estimate of $B_3$ for the 3D piece of ice ($\approx 0.85\ k_B T_0$ near $T_0 = 273$ K) follows from the free energy of the ice/water interface, $\approx 32$ erg/cm$^2$ [10], and, because an H$_2$O molecule occupies $\approx 10$Å$^2$ of the interface, $B_3 \approx 320 \times 10^{-16}$ erg $\approx 1.9$ kJ/mol, or $B_3 \approx 0.85\ k_B T_0$ per one surface H$_2$O molecule. Thus,

$$\frac{B_3}{-\Delta\mu} \approx \frac{85°}{\Delta T}, \tag{6}$$

and the transition state free energy for the 3D ice crystal formation can be estimated as

$$\frac{G_{3D}^\#}{k_B T_0} \approx 12 \left(\frac{100°}{\Delta T}\right)^2. \tag{7}$$

When $\Delta T$ is small, $G_{3D}^\#$ is extremely high: at $\Delta T = 1°$, $G_{3D}^\# \approx 300\,000$ kJ/mol (which is at least 1000 times higher than the enthalpy of interaction of non-polymeric ions and molecules with their environment [11, 12]); at $\Delta T = 10°$, $G_{3D}^\# \approx 3\,000$ kJ/mol, and only at $\Delta T = 30°$, $G_{3D}^\# \approx 300$ kJ/mol, which approaches the enthalpy of interaction of small molecules and ions with their environment.

**Table 1 | The time of homogeneous ice nucleation in bulk water in the absence of ice-binding surfaces**

| Volume (V) | Waters in the volume, $N_V$ | $TIME_{3D}(\Delta T)$ | | | | |
|---|---|---|---|---|---|---|
| | | $-10°C$: $\Delta T = 10°$ | $-30°C$: $\Delta T = 30°$ | $-35°C$: $\Delta T = 35°$ | $-40°C$: $\Delta T = 40°$ | $-50°C^*$: $\Delta T = 50°$ |
| Lake (30 km$^3$=3·10$^{16}$ cm$^3$) | $10^{39}$ | $10^{467}$ years | $10^4$ years | 0.001 seconds | $\geq 10^{13}$ nucleations/s | $\geq 10^{25}$ nucleations/s |
| Mug 3·10$^2$ cm$^3$ | $10^{25}$ | $10^{481}$ years | $10^{18}$ years | 100 years | 10 seconds | $\geq 10^{11}$ nucleations/s$^*$ |
| Bacterium (3 μm$^3$=3·10$^{-12}$ cm$^3$) | $10^{11}$ | $10^{504}$ years | $10^{41}$ years | $10^{16}$ years | $10^6$ years | 10 minutes |

Footnotes:
[*]"... at fifty below spittle crackled on the snow ..." [Jack London. "To build a fire"]
[**]i.e., 1 nucleation per second per $10^{26}$ waters (3 liters)
[***]i.e., 1 nucleation per second per $10^{14}$ waters (3000 μm$^3$)

If the volume contains $N_V$ waters, and a nucleus can arise around any of them, the characteristic time of appearance of one and only 3D ice nucleus in this volume can be estimated as

$$TIME_{3D} \sim \tau \cdot \exp\left(\frac{G_{3D}^{\#}}{k_B T}\right)/N_V \sim \frac{10^{-5}(1°/\Delta T)}{N_V} \cdot \exp\left(12\left(\frac{100°}{\Delta T}\right)^2\right) \text{ seconds.} \tag{8}$$

The characteristic time of homogeneous (occurring in bulk water) initiation of ice formation at different temperatures in vessels of different volumes is illustrated in Table 1.

If ice forms not around a water molecule, but around some other, "foreign" small molecule or ion, which attracts ice very strongly, then its nucleation time dramatically decreases. But a decrease by even 50 orders of magnitude (which corresponds to the practically maximal, ~300 kJ/mol [11, 12], attraction of a "foreign" molecule to its environment) may manifest itself in observable phenomena only at temperatures below $-20° - -25°C$.

Concluding this part, it is worthwhile to note that the resulting diameter of the ice "seed" corresponds (see Eqs. (2a), (6)) to the length of a row of ≈60 waters (≈200Å) at $-10°C$, and ≈40Å at $-50°C$. These estimates, as well as those in Table 1, are, of course, rather approximate when $\Delta T$ is not small.

*The time of ice growth after the nucleation*

So far, we have taken into account only the time of initiation of ice formation but not the time of its growth. The reason is that growth is a relatively fast process. Since after the formation of the ice nucleus the remaining H₂O molecules attach to it more or less independently, the $\tau$ value is both the time of attachment of one H₂O molecule to the ice and the time of growth of one layer of H₂O molecules on the ice. Thus, the time of addition of one new layer of aqueous molecules to the ice at temperatures below $-10°C$ is less than $10^{-6}$ s (see Eq. (5)). So, in 0.01 seconds, 10 000 new layers of H₂O molecules join the ice (this new layer is ~3 μm thick, like a bacterium), and the ice will be ~30 mm thick (the radius of a mug) within minutes, – while when speaking on the ice initiation at $-30°C$ and above, I spoke about much longer times. Thus, the time of ice growth can be neglected as compared to the time of its appearance.

So,
1) the time of homogeneous nucleation of ice soars to infinity when the temperature approaches $0°C$;
2) ice can never arise by homogeneous nucleation in bulk water at temperatures above $-30°C$, which means that above $-30°C$ one cannot use a homogeneous model of the initiation of freezing that starts with the ice formation in bulk water. However, this model may be applicable at lower temperatures when (see Eq. (4)) $-\Delta\mu \geq 0.4 k_B T_0$ (cf. [1], ch. 14, §4; [2], ch. 5.3). According to [13], freezing of aqueous droplets in the atmosphere occurs at -35°C and below, and the maximum rate of their freezing is observed at -42° — -46°C.

It's another matter when ice arises not in bulk water, but on its surface, i.e., by the heterogeneous [1-4] nucleation (see Table 2 below).

Any surface, even that of a dust particle is suitable for this, if only it would attract ice much stronger than water.

*The heterogeneous nucleation: ice formation at a surface*

So, let us now consider the initiation of ice by a heterogeneous nucleation, i.e., that occurring on a surface of water [1-4] (Fig. 3). The surface that I have in mind speaking on the ice nucleation, is an ice surface because when ice is not attracted by the surface (or attracted less than liquid water), it does not grow on it at all; and when its adhesion to the surface is stronger than that of water, ice will cover this surface with an initial monomolecular ice layer even at temperatures above 0°C (but will not grow further at these temperatures), and the massive ice growth (which is of interest for us) only occurs on that icy surface at temperatures below 0°C.

The free energy of an arising monomolecular ice layer that grows upon the ice surface is determined only by the perimeter of this new layer since the emerging surface energy of the upper (in Fig. 3) side of the new layer is compensated by the disappearance of the surface energy of the "old" ice surface that is now covered by the new layer. According to Eq. (3), the activation free energy is determined by the perimeter of a nucleating layer as $G_{2D}^{\#} \approx 4B_2\left(\frac{B_2}{-\Delta\mu}\right)$, where $B_2$, the additional free energy of a perimeter's molecule, depends on the interaction of molecules within the new layer (see above). Figure 3 shows that the value of $B_2$ and, in general, the presence of the activation barrier itself depends on the structure of the surface onto which a new layer of ice grows.

We shall start with examining the initiation of ice growth on a smooth flat surface (Fig. 3A).

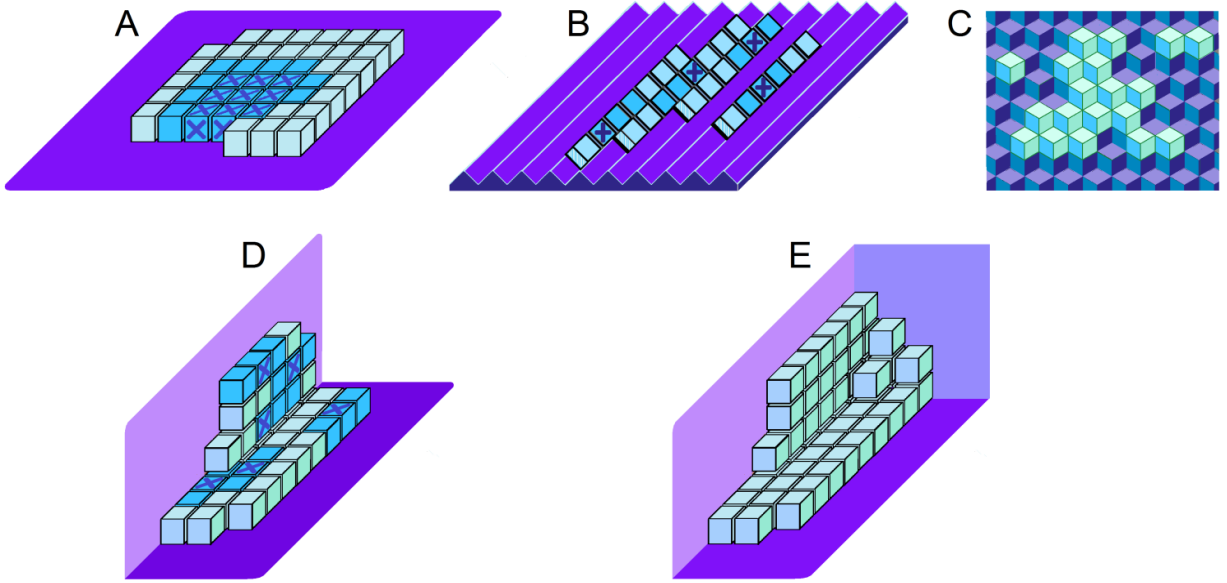

FIG. 3. Scheme of ice layers on ice-binding surfaces (shown in blue-violet). Other designations are the same as in Fig. 1. The plain surface, to which a new layer of ice binds, can, in principle, be either flat and smooth (A), or corrugated (B) or fluted (C). The molecules of the layer formed on a smooth (A) surface interact in the same way as in each layer of a 3D body (see Fig. 1A), so the additional free energy $B_2$ of a perimeter molecule of such a layer ($B_{\text{2D\_smooth}}$) is significant and approaches the free energy $B_3$ of a molecule on the surface of the 3D body. Because of that, the ice nucleation at a smooth surface also (as in bulk water) can only occur at sub-zero temperatures and is slow. In the layer formed on the corrugated (B) surface, the molecules strongly interact along the grooves, but weakly interact across the grooves; therefore, in this case, there is no single value of $B_2$; instead there is a significant, approaching $B_3$, value $B_2^+$ for the layer sides that are perpendicular to the grooves, and a small value $B_2^-$ for those that are parallel to the grooves. In the layer formed on the fluted (C) surface, all molecules of the new layer weakly interact with one another, yielding a small $B_2 \approx B_2^-$. Weak interactions of molecules within new layers shown in panels (B) and (C) provide a rapid initiation of ice at the corresponding surfaces. (D, E): The once-bent (D) and twice-bent (E) surfaces that facilitate the rapid initiation of ice based on such surfaces. A rapid ice formation upon grooved surfaces shown in panels B, C, D, E can occur even at exactly $0°C$ (see the text).

In this case, $B_{\text{2D\_smooth}}$, the additional free energy of a perimeter molecule of the smooth layer, should approach $B_3$, the additional free energy of a molecule at the boundary of the body shown in Fig. 1A (although it still seems that $B_2$ here should be a little smaller than $B_3$ due to the attraction of edge molecules of the new layer to the underlying surface (Fig. 3A)).

Anyhow, for simplicity of the following calculations, I assume that $B_{\text{2D\_smooth}} \approx B_3 \approx 0.85\ k_B T_0$.

Using Eqs. (2), (2a), (4), (6), we can estimate the diameter of the ice "nuclei" and "seeds" formed at smooth surfaces; they are inversely proportional to the value of $\Delta T$ and approximately correspond to the length of a row of ≈70 waters (≈200Å) at $-2.5°C$, of ≈35 waters (≈100Å) at $-5°C$, and ≈60Å at $-9°C$ for a nucleus, and to ≈400Å at $-2.5°C$, ≈200Å at $-5°C$, and ≈110Å at $-9°C$ for a seed.

The free energy of the nucleus at a smooth surface is estimated using Eqs. (3), (4), (6):

$$\frac{G^{\#}_{\text{2D\_smooth}}}{k_B T_0} \approx \frac{300°}{\Delta T}. \tag{9}$$

When $\Delta T$ is small, $G^{\#}_{\text{2D\_smooth}}$ is high: at $\Delta T = 0.5°$, $G^{\#}_{\text{2D\_smooth}} \approx 1400$ kJ/mol (which is 10 times higher than the enthalpy of the interaction of non-polymeric ions and molecules with their environment [10, 11]); and only at $\Delta T = 3°$, $G^{\#}_{2D\_smooth} \approx 200$ kJ/mol, which already approaches the maximal enthalpy of the interaction of non-polymeric molecules and ions with their environment.

Thus, the characteristic time of appearance of one 2D ice nucleus on a smooth surface covered by $N_S$ waters can be estimated as

$$TIME_{2D\_smooth} \sim \tau \cdot \exp\left(\frac{G^{\#}_{2D\_smooth}}{k_B T}\right)/N_S \sim \frac{10^{-5(1°/\Delta T)}}{N_S} \cdot \exp\left(\frac{300°}{\Delta T}\right) \text{ seconds}; \quad (10)$$

here, we again use Eq.(5) to estimate $\tau$.

The characteristic times of heterogeneous ice nucleation at flat and smooth surfaces in vessels with various surface areas are presented in Table 2 for different temperatures.

Because the number of aqueous molecules on the surface, $N_S$, is not uniquely determined by the number of molecules in the volume, $N_V$, we can use two natural limits, $N_{S\_max}=N_V$ (all waters are in contact with ice-binding surfaces, i.e., with all walls of the vessel, with all protrusions and depressions of the walls, with all surfaces of dust-like particles floating in the liquid, etc.), and $N_{S\_min}\approx(N_V)^{2/3}$ (one flat inner surface of the vessel only is the ice-binding surface, and the liquid contains no dust or other inclusions), and their average value $N_{S\_av}\approx(N_{S\_min}\times N_{S\_min})^{1/2}$ approximately corresponding to the presence of some dust in the water.

If ice forms at a smooth surface around some extraneous impurity (non-aqueous molecule or ion) that strongly attracts ice, then the additional stabilization of the nucleus by this impurity can accelerate ice nucleation (although the number of such "impurity nucleation centers" is much lower than the number of waters on the wall), but (see above), this effect can be observed only at temperatures below $-2°C$.

**Table 2 | The time of heterogeneous ice nucleation at flat and smooth ice-binding surfaces**

| Volume (V) | Number of waters | | $TIME_{2D\_flat}(\Delta T)$ | | | | | |
|---|---|---|---|---|---|---|---|---|
| | in the volume, $N_V$ | on the surface, $N_S$ | $-2.5°C^{\dagger}$: $\Delta T=2.5°$ | $-3°C^{\dagger}$: $\Delta T=3°$ | $-4°C^{\dagger}$: $\Delta T=4°$ | $-5°C^{\dagger}$: $\Delta T=5°$ | $-7°C^{\dagger}$: $\Delta T=7°$ | $-9°C^{\dagger}$: $\Delta T=9°$ |
| Lake (30 km³) | $10^{39}$ | $N_{S\_max}=10^{39}$ | 4 months | 0.1 sec. | $10^{-12}$ sec. | $10^6$-$10^{19}$ nucleations per second | $10^{13}$-$10^{26}$ nucleations per second | $10^{17}$-$10^{30}$ nucleations per second |
| | | $N_{S\_av}\approx 10^{32}$ | **$10^6$ years** | **1 day** | **$10^{-6}$ sec.** | | | |
| | | $N_{S\_min}\approx 10^{26}$ | $10^{12}$ years | 3000 years | 10 sec. | | | |
| Puddle 300 dm³ | $10^{28}$ | $N_{S\_max}=10^{28}$ | $10^{10}$ years | 30 years | 0.1 sec | $10^{-8}$ sec. | $10^7$-$10^{15}$ nucleations per second | $10^{10}$-$10^{19}$ nucleations per second |
| | | $N_{S\_av}\approx 10^{23}$ | **$10^{15}$ years** | **$10^6$ years** | **1 hour** | **0.01 sec.** | | |
| | | $N_{S\_min}\approx 10^{19}$ | $10^{19}$ years | $10^{10}$ years | 1 month | 5 min. | | |
| Mug 300 cm³ | $10^{25}$ | $N_{S\_max}=10^{25}$ | $10^{13}$ years | $3\cdot 10^4$ years | 2 minutes | $10^{-5}$ sec. | $10^4$-$10^{12}$ nucleations per second | $10^8$-$10^{16}$ nucleations per second |
| | | $N_{S\_av}\approx 10^{21}$ | **$10^{17}$ years** | **$10^8$ years** | **3 days** | **1 sec.** | | |
| | | $N_{S\_min}\approx 10^{16}$ | $10^{22}$ years | $10^{13}$ years | 100 years | 3 hours | | |
| Bacterium (3 μm³) | $10^{11}$ | $N_{S\_max}=10^{11}$ | $10^{27}$ years | $10^{18}$ years | $10^7$ years | 10 years | 1 minute. | 0.01 sec. |
| | | $N_{S\_av}\approx 10^9$ | **$10^{29}$ years** | **$10^{20}$ years** | **$10^9$ years** | **500 years** | **1 hour** | **0.1 sec.** |
| | | $N_{S\_min}\approx 10^8$ | $10^{31}$ years | $10^{22}$ years | $10^{11}$ years | $3\cdot 10^4$ years | 2 days | 10 sec. |

Footnotes:

$^{\dagger}$ The data are tabulated under the assumption that $B_2 = B_3 = 0.85\, k_B T_0$, although (see text) $B_2$ is apparently a little smaller than $B_3$, and then the time estimates given in the Table are slightly overestimated (and the temperatures, accordingly, are slightly underestimated). Incorporation of a separate ice-attracting non-aqueous molecule or ion in the surface where ice grows may additionally raise the temperatures given in the Table by $\approx +1°$.

$^*$ i.e., 1 nucleation per second per $10^{20}$ surface waters (10 m²).

$^{**}$ i.e., 1 nucleation per second per $10^{13}$ surface waters (1 mm²).

$^{***}$ i.e., 1 nucleation per second per $10^9$ surface waters (100 μm²).

So,
1) the time of heterogeneous ice nucleation at the smooth surface is also very temperature-dependent, although not as much as that of the homogeneous one, and also tends to infinity when the temperature approaches 0°C;
2) at small sub-zero temperatures, ice cannot arise by heterogeneous nucleation on smooth walls, even when they attract ice: in macroscopic vessels it cannot arise at temperatures above -2⁰— -3⁰C, in microscopic vessels (animal and plant cells, bacteria) it cannot arise at temperatures above -6⁰— -7⁰C.

This means that the conventional model of freezing, that implies ice initiation on the aqueous surface, or on a smooth wall, or on smooth dust particles, also cannot *per se* explain freezing of a puddle, or of a cup of water, or of a bacteria at 0°C: the above analysis shows that freezing can begin only at a substantially negative temperature, below $\approx -3°C$ (or $\approx -2°C$ with additional stabilization of the ice nucleus by an ice-initiating "wall impurity"), but not at $0°C$.

**Factors that can dramatically accelerate heterogeneous ice nucleation at ice-binding surfaces**

However, our daily experience (leaving bacteria aside for now) indicates that macroscopic vessels – such as puddles – freeze at practically $0°C$ (not at $-3°C$ or $-2°C$). How can this be?
In fact, the question splits into two:
(1) Can water acquire a substantially negative temperature at zero ambient temperature?
(2) Is it possible to form ice in a way that avoids the formation of such a "nucleus" of a new phase that has infinitely high surface free energy at $0°C$?
Let's start by answering the first question.

*Decrease in water temperature due to evaporation accelerates freezing*

In an open vessel (lake, puddle, cup), cooling of the surface water is possible due to evaporation; it will be shown below that the surface water can be cooled to ≈-4°C at 0°C in the environment.

Water cooling by evaporation is used in devices such as a psychrometric hygrometer, which measures air humidity by comparing the temperature of a dry thermometer and the one covered with wet-cloth [14, 15]. A typical result here: +18.5°C on the wet thermometer, +22°C on the dry one at the relative humidity of 70%; or +2°C on the wet thermometer, +4°C on the dry one at the relative humidity of 70%; or +0.6°C on the wet thermometer, + 4°C on the dry one at the relative humidity of 50%; or -2.85°C on the wet (more precisely, on an already frozen) thermometer, 0°C on the dry one at the relative humidity of 50%.

Reduced, as a result of intense evaporation, the temperature of the surface layer of water is observed in freezing natural reservoirs [16]. This is enough for ice formation on the dust particles present in water, but NOT enough to initiate ice without them [16] (see also [13]). (Note, however, that this way of lowering the temperature is NOT applicable to closed vessels, such as intact cells that we shall consider later).

The expected decrease in the temperature of the reservoir's surface layer due to evaporation can be estimated as follows.
1) Considering water, which has the surface temperature $T_{ws}$, and air, having temperature $T_0$ and relative humidity $H_\%$ (measured in %), one can estimate the water evaporation rate $E_{mass}$ (measured in *mm/day*) using the Penman-Shuttleworth equation [17-20], which, for the open water, in the absence of solar irradiation and wind, reads as

$$\left[\frac{E_{\text{mass}}}{mm/day}\right] = 6.43 \frac{\left[\frac{\delta p}{kPa}\right]}{\left[\frac{\lambda_v}{MJ/kg}\right]\left(1+\frac{m}{\gamma}\right)}. \tag{11}$$

Here, $\lambda_v$ is the latent heat of water vaporization (in $MJ/kg$): $\approx 2.5\ MJ/kg$ at $0°C$ [9]; $m$ is the slope of the saturation vapor pressure curve (in *Pa/K*): ≈44 *Pa/K* at $0°C$, $p_{\text{air}}$=1 atm [18, 19]; $\gamma$ is the psychrometric constant (in *Pa/K*): ≈66 *Pa/K* at $0°C$, $p_{\text{air}}$=1 atm [18, 19], and $\delta p$ is the water vapor pressure deficit (in $kPa$): $\delta p = p_s(T_{ws}) - p_v(T_0, H_\%)$, where $p_s(T_{ws})$ is the saturated vapor pressure (in $kPa$) at temperature $T_{ws}$ of the water surface, while $p_v(T_0, H_\%)$ is the vapor pressure (in $kPa$) in the air having temperature $T_0$ ($T_0 \approx 273°K$ in this work) and relative humidity $H_\%$ (in %). As known [14, 20], $p_v(T_0, H_\%) = p_s(T_0) \cdot \frac{H_\%}{100\%}$, and the value of $p_s(T_0=273°K)$ is $\approx 0.61$ $kPa$ [8, 21].

The water vapor dew-point temperature $T_{dp}(T_0, H_\%)$ (tabulated in [14, 22]) satisfies a condition $p_s(T_{dp}(T_0, H_\%)) = p_v(T_0, H_\%)$ [14, 20], so that $p_s(T_{dp}(T_0, H_\%)) = p_s(T_0) \cdot \frac{H_\%}{100\%}$. As a result,

$$\delta p = p_s(T_{ws}) - p_v(T_0, H_\%) = p_s(T_{ws}) - p_s(T_{dp}(T_0, H_\%)) \approx \left.\frac{dp_s}{dT}\right|_{T_0} \cdot [T_{ws} - T_{dp}(T_0, H_\%)] \approx$$
$$\left.\frac{dp_s}{dT}\right|_{T_0} \cdot [T_0 - T_{dp}(T_0, H_\%)] \cdot \frac{T_{ws}-T_{dp}(T_0,H_\%)}{T_0-T_{dp}(T_0,H_\%)} \approx [p_s(T_0) - p_s(T_{dp}(T_0, H_\%))] \cdot \frac{T_{ws}-T_{dp}(T_0,H_\%)}{T_0-T_{dp}(T_0,H_\%)} =$$
$$\left[p_s(T_0) - p_s(T_0) \cdot \frac{H_\%}{100\%}\right] \cdot \frac{T_{ws}-T_{dp}(T_0,H_\%)}{T_0-T_{dp}(T_0,H_\%)} = p_s(T_0) \cdot \frac{100\%-H_\%}{100\%} \cdot \frac{T_{ws}-T_{dp}(T_0,H_\%)}{T_0-T_{dp}(T_0,H_\%)}. \tag{12}$$

Numerically (at the air temperature $T_0 \approx 273°K$),

$$\left[\frac{E_{mass}}{mm/day}\right] = 6.43 \frac{\left[\frac{p_s(T_0)}{kPa}\right] \cdot \frac{100\%-H_\%}{100\%}}{\left[\frac{\lambda_v}{MJ/kg}\right]\left(1+\frac{m}{\gamma}\right)} \cdot \frac{T_{ws}-T_{dp}(T_0,H_\%)}{T_0-T_{dp}(T_0,H_\%)} = 6.43 \frac{[0.61] \cdot \frac{100\%-H_\%}{100\%}}{[2.50](1.67)} \cdot \frac{T_{ws}-T_{dp}(T_0,H_\%)}{T_0-T_{dp}(T_0,H_\%)} \approx$$
$$\left[\frac{1.}{mm/day}\right] \cdot \frac{100\%-H_\%}{100\%} \cdot \frac{T_{ws}-T_{dp}(T_0,H_\%)}{T_0-T_{dp}(T_0,H_\%)}. \tag{13}$$

This means that open water evaporates at a rate of about 1 mm per day when its surface has the same temperature $T_0$ (i.e., 0°C) as the absolutely dry air above it; at a rate of $\approx 0.5$ mm per day when the air's humidity is $\approx 50\%$; and of $\approx 0.3$ mm per day when the air's humidity is close to 70%.

I obtained more or less close numbers doing probing experiments with a layer of water on my windowsill near Moscow in March, at about 0° temperature and humidity of 50 to 70%. A somewhat higher evaporability was recorded by meteorologists [23] (on average throughout the Moscow region, mostly covered by vegetation, and only to a small extent by open water) for those months when the temperature here is close to 0°: about 0.4 mm per day in November, when the average humidity was $\approx 76\%$, and about 0.7 mm per day in March, when the average humidity was $\approx 65\%$. It seems that this increase in evaporability in natural conditions of Moscow region is due to the action of wind and solar irradiation [17-20], which have been now ignored in a simplified Eq. (11) and absent from my on-windowsill experiments.

In the absence of wind and solar irradiation, evaporation of a millimeter layer, i.e., of 1 kg of water from 1 m$^2$, absorbs 2.5 MJ [9], and if this occurs in 1 day, i.e., 86400 seconds, then the density of the heat flux required for the evaporation of this layer is

$$J_{T_0,0\%} \approx 30 \text{ W/m}^2. \tag{14}$$

Such a heat flux density would correspond to zero humidity of the air with a temperature of $T_0 \approx 273°K$ and the same, $T_{ws} \approx 273°K$, temperature of the water surface. If the air with temperature 0°C has a relative humidity of $H_\%$, and the temperature of the water surface is $T_{ws}$, then the density of the heat flux supporting the evaporation is

$$J_{T_{ws},H_\%} = J_{T_0,0\%} \cdot \frac{100\%-H_\%}{100\%} \cdot \frac{T_{ws}-T_{dp}(T_0,H_\%)}{T_0-T_{dp}(T_0,H_\%)}. \tag{15}$$

Let us return to a decrease in the temperature of open water (in a mug, puddle, lake, etc.) upon evaporation from its surface at the air temperature of about 0°C.

Formally, the problem is posed as follows:

Consider a one-dimensional homogeneous system of length $L$ (Fig. 4) without internal heat sources, but with heat outflow from its "evaporating" surface (in Fig. 4 – the left ($x = 0$) side). Note that when considering evaporation at a temperature of about 0°C, convection can be neglected [24], since the water cooled by evaporation (and therefore, less dense at temperatures below +4°C) does not go down from the evaporating water surface, and the air cooled by evaporation and therefore more dense does not rise up from it.

The temperature $\theta(x, t)$ in this system depends on time ($t \geq 0$) and place ($0 \leq x \leq L$). The density of the heat flux in the system, $J(x,t)$, obeys the Fourier's law, $J(x,t) = -\kappa \cdot \frac{\partial \theta(x,t)}{\partial x}$, where $\kappa$ is the thermal conductivity coefficient. The density of the time-dependent heat flux from the system is $J(0,t) = J_{T_{ws}(t)}$, where $T_{ws}(t) = \theta(0,t)$.

The divergence of the heat flux density determines the temperature change: $\frac{\partial J(x,t)}{\partial x} = -C_1 \cdot \frac{\partial \theta(x,t)}{\partial t}$ (where $C_1$ is the heat capacity per unit volume). Combining the above two expressions, we have a classical homogeneous Laplace equation for thermal conductivity [25]:

$$\frac{\partial \theta(x,t)}{\partial t} = \frac{\kappa}{C_1} \cdot \frac{\partial^2 \theta(x,t)}{\partial x^2}. \tag{16}$$

The initial (at $t = 0$) condition for this equation: the same temperature for all $0 \leq x \leq L$:

$$\theta(x,0) = T_0 = const \tag{17}$$

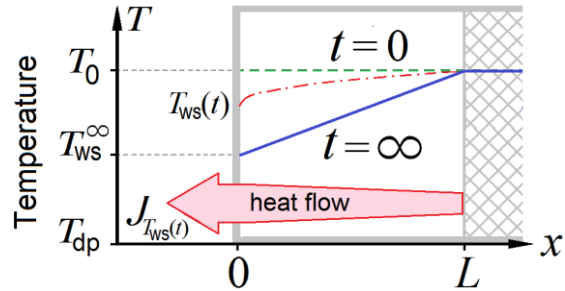

FIG. 4. Schematic presentation of a one-dimensional system with the heat flux of $J_{T_{ws}(t)}$ density coming out at $x = 0$, and a constant temperature of $T_0$ at $x = L$. The green dashed line is the profile of temperature $\theta(x, 0) = T_0$ at the beginning of the process (at $t = 0$); the solid blue line is the profile of temperature $\theta(x, \infty)$ in the stationary regime (at $t = \infty$); the dashed/dotted red line is a profile of temperature $\theta(x, t)$ in the intermediate regime (the temperature $T_{ws}(t)$ at $x = 0$ and the heat flux shown by the red arrow increasing to $x = 0$ refer to the intermediate time $t$). $T_{dp}$ is the "dew point" temperature (see text).

The boundary condition at $x = L$ ("constant temperature end") and $t \geq 0$ has an evident inhomogeneous form

$$\theta(L,t) = T_0 = const. \tag{18}$$

The boundary condition at $x = 0$, $t \geq 0$: The density of a heat flux out of the system (Fig. 4) at this "radiating end" is $J_{T_{ws}(t)} = -J_0 \frac{T_{ws}(t) - T_{dp}}{T_0 - T_{dp}}$, where (see Eq. (15)): $J_0 \equiv J_{T_0, 0\%} \cdot \frac{100\% - H_\%}{100\%} = const$ is the flux density at $\theta(0,0) = T_0$; $T_{dp} \equiv T_{dp}(T_0, H_\%) = const < T_0$ is the dew point temperature at a given air relative humidity $H_\% = const$ and air temperature $T_0 = const$; and $T_{ws}(t) = \theta(0,t)$; that is, the boundary condition at this system's end has an inhomogeneous form

$$\theta(0,t) = -J_{T_{ws}(t)} \frac{T_0 - T_{dp}}{J_0} + T_{dp}, \tag{19}$$

where

$$J_{T_{ws}(t)} = -\kappa \cdot \frac{\partial \theta}{\partial x}(x,t)\bigg|_{x \to 0}. \tag{20}$$

Numerical parameters: $T_0 \approx 273°K$ (i.e., $0°C$); $\kappa \approx 0.6$ W/(m·K) at $0°C$ [9]; $C_1$ = (specific heat capacity of water at $0°C$) × (its density at $0°C$) $\approx 4.2$ MJ/(m$^3$·K) [9] (thus, $\frac{\kappa}{C_1} \approx 0.14 \cdot 10^{-6}$ m$^2$/s); $J_0 = (30\,W/m^2) \cdot \frac{100\% - H_\%}{100\%}$ at $0°C$ (see Eqs. (14), (15)); the values $T_{dp}(T_0, H_\%)$ are tabulated in [14, 22].

The general solution of Eq. (16) with the initial and boundary conditions given by Equations (17), (18), (19) (that is, the solution that describes the entire spectrum of relaxation times and describes the entire change in the temperature distribution with time) can be achieved by methods that are standard for mathematical physics (such as separation of variables, calculation of coefficients of infinite Fourier series and summation of these series) but rather voluminous [25, 26]. I take the liberty not to reproduce them here but focus on a simplified solution with a clear physical sense that answers only the question of interest to us – by how much, ultimately, the evaporation will lower the water temperature. To this end, we have only to find a steady state solution of the above equations.

At the beginning of the process, until the water surface temperature ($T_{ws}$) is close to the temperature of the medium ($T_0$, which corresponds to 0°C), the density of the heat flow needed for the evaporation at a typical air humidity is $J_{T_0,70\%} \approx 9$ W/m² at 70% humidity and $J_{T_0,80\%} \approx 6$ W/m² at 80% humidity according to Eq. (15).

Almost all of this heat is absorbed from water, whose thermal conductivity ($\kappa \approx 0.6$ W/(m·K) at 0°C [9]) is much greater than that of air [27, 28]. As it has been already mentioned, one can neglect convection when considering evaporation at about 0°C [24].

Therefore, according to the Fourier's law of thermal conductivity [9], the initial near-surface temperature gradient of the heat fluxes that supports the evaporation are: $\text{grad}(T)|_{ws} = -J_{T_0,70\%}/\kappa \approx$ -15 K/m and $J_{T_0,80\%}/\kappa \approx 10$ K/m at the above mentioned initial temperature $T_{ws} = T_0$ of the water surface and the typical 70% – 80% humidity.

As water evaporates, the temperature $T_{ws}$ of its surface drops (Fig. 4) and, according to Eq. (15), the heat flux density decreases from $J_{T_0,H\%}$ to $J_{T_{ws},H\%} = J_{T_0,H\%} \cdot \frac{T_{ws}-T_{dp}(T_0,H\%)}{T_0-T_{dp}(T_0,H\%)}$, where $T_{dp} < T_0$ is the dew point temperature [9].

Let us now consider the steady-state regime at $t \to \infty$, when the temperature ceases to change with time and uniformly decreases from the temperature $T_0$ maintained at a depth $L$ under the surface to the surface temperature $T_{ws}^\infty$. Then the constant (along the heat flow going from the bottom to the surface) temperature gradient is $\frac{T_{ws}^\infty-T_0}{L}$, and the heat flux density is $J_{T_0,H\%} \cdot \frac{T_{ws}^\infty-T_{dp}(T_0,H\%)}{T_0-T_{dp}(T_0,H\%)}$, so that

$$-\left[J_{T_0,H\%} \cdot \frac{T_{ws}^\infty-T_{dp}(T_0,H\%)}{T_0-T_{dp}(T_0,H\%)}\right]/\kappa = \frac{T_{ws}^\infty-T_0}{L}, \tag{21}$$

from where

$$T_{ws}^\infty = T_0 - \frac{\left(\frac{J_{T_0,H\%}}{\kappa} \cdot L\right) \times [T_0-T_{dp}(T_0,H\%)]}{\left(\frac{J_{T_0,H\%}}{\kappa} \cdot L\right)+[T_0-T_{dp}(T_0,H\%)]}. \tag{22}$$

For a large $L$ value, the surface temperature $T_{ws}^\infty$ tends to $T_{dp}(T_0,H\%)$, i.e., at a typical humidity of 70-80% and the ambient temperature $T_0$, corresponding to 0°C, $T_{ws}^\infty$ tends to -4.9°C — -3.1°C [14, 22] (Note that the physical depth of the reservoir usually does not matter much in this effect, since the thermal conductivity of the soil is, as a rule, rather close to the thermal conductivity of water [9, 27, 28], and therefore the depth $L$, at which the temperature remains close to $T_0$, is unlimited, so that the surface temperature $T_{ws}^\infty$ normally approaches $T_{dp}(T_0,H\%)$.) And in water supercooled to $T_{dp}(T_0,H\%)$ = -4.9°C — -3.1°C, ice forms quickly (see Table 2 and its first footnote), although it cannot form at all above -2.5°C (see Table 2).

If the depth $L$ is small, and a thermally insulating layer (wood, insulating bricks, etc. [27, 28]) lies under it, the evaporation will eventually bring all the water, down to the very bottom, to the same dew point temperature, $T_{dp}(T_0,H\%)$, – and again the freezing will be fast.

It's another matter when under a thin layer of water lies a layer of high thermal conductivity (for example, a metal) that maintains the temperature $T_0$ at the small depth $L$, such that $\left(\frac{J_{T_0,H\%}}{\kappa} \cdot L\right) <$ $[T_0 - T_{dp}(T_0,H\%)]$ (e.g., $L$ is less than ≈30 cm at a humidity of 70-80% and 0°C): then the water temperature will not drop below $[T_0 + T_{dp}(T_0,H\%)]/2$, that is, below -2.45°C — -1.55 °C, and ice will be not able to form.

The time required to set a stationary regime is approximately estimated as follows:

If in a layer of thickness $L$ the temperature, in a stationary regime, varies linearly from $T_0$ to $T_{ws}^\infty$ (see Fig. 4), the average layer temperature is $\frac{T_0+T_{ws}^\infty}{2}$, and thus the thermal energy in this layer (of area

S) changes (as compared to that at $T \equiv T_0$) by

$$\Delta Q = SL \cdot C_1 \cdot \left[\frac{T_{ws}^\infty + T_0}{2} - T_0\right] = -SL \cdot C_1 \cdot \frac{T_0 - T_{ws}^\infty}{2}, \quad (23)$$

where $C_1$ is the heat capacity per unit volume of water at $T = T_0$, i.e., $C_1$ = (specific heat of water at $T = T_0$) × (its density at $T = T_0$) = 4.2 MJ/(m³·K) [9].

For the given $\text{grad}(T) = \frac{T_0 - T_{ws}^\infty}{L}$, the heat flux density is $J = -\kappa \cdot \text{grad}(T)$, where $\kappa_{water} \approx 0.6$ W/(m·K) [9], i.e., the amount of heat that flowed out during the time $t$ is

$$Q_t = S \cdot J \cdot t = -S \cdot \kappa \cdot \frac{T_0 - T_{ws}^\infty}{L} \cdot t. \quad (24)$$

The flow will be set when $Q_t \sim \Delta Q$, i.e $-S \cdot \kappa \cdot \frac{T_0 - T_{ws}^\infty}{L} \cdot t \sim -SL \cdot C_1 \cdot \frac{T_0 - T_{ws}^\infty}{2}$, or

$$t \sim L^2 \cdot \frac{C_1}{2\kappa} = \left(\frac{L}{m}\right)^2 \cdot m^2 \cdot 4.2 \frac{\left[\frac{MJ}{m^3 \cdot K}\right]}{2 \cdot 0.6 \frac{W}{m \cdot K}} \cong \left(\frac{L}{m}\right)^2 3.5 \times 10^6 \, s. \quad (25)$$

Therefore, at $L = 10$ cm, — $t \sim 35000$ seconds (i.e., in a puddle the heat flux will be set overnight), but in a reservoir of one-meter depth, this will take a month.

It makes sense to add here that a strict general mathematical solution of Equation (16), with boundary conditions outlined by Eqs. (18), (19), gives the following spectrum of exponential relaxation times: $\frac{C_1 L^2}{\kappa}\left(\frac{2}{\pi}\right)^2$, $\frac{C_1 L^2}{\kappa}\left(\frac{2}{3\pi}\right)^2$, $\frac{C_1 L^2}{\kappa}\left(\frac{2}{5\pi}\right)^2$, $\frac{C_1 L^2}{\kappa}\left(\frac{2}{7\pi}\right)^2$, ... The largest of them, $\frac{C_1 L^2}{\kappa}\left(\frac{2}{\pi}\right)^2 \approx \frac{C_1 L^2}{2.47\kappa}$, is close to the value $t \sim \frac{C_1 L^2}{2\kappa}$ obtained from simple physical considerations in equation (25). It corresponds to the time of approaching the asymptotic stationary solution $\theta(x, t \to \infty)$, where the temperature decreases linearly (see Fig. 4) in the interval $L \geq x \geq 0$ from $\theta(L, t \to \infty) = T_0$ to $\theta(0, t \to \infty) = T_{ws}^\infty$, with the same surface temperature $T_{ws}^\infty$, which was obtained in Equation (22).

*Accelerating the initiation of ice formation by grooved surfaces*

Now we can ask ourselves another question – is there a possibility of ice formation by a pathway avoiding the formation of a compact "nucleus" of the new phase, whose surface energy makes the initiation of ice formation possible only at substantially negative temperatures?

In principle, this is possible – if the ice-binding surface is grooved – "corrugated" (Fig. 3B), or "fluted" (Fig. 3C), or bent once (Fig. 3D) or twice (Fig. 3E) in such a way that a new water molecule can almost always (Fig. 3D) or even always (Fig. 3E) be added to ice without increasing the ice surface. (It goes without saying that Figure 3 demonstrates only the basic schematic presentations of such surfaces rather than detailed ice structures.)

The "corrugated" and especially the "fluted" surfaces (and then the ice that has grown on them) do not allow many strong contacts (Fig. 3B) or allow no contacts at all (Fig. 3C) between molecules in the nascent layer. At a "corrugated" surface, the new coming molecules interact weakly with molecules of the nascent layer positioned in the neighboring grooves (although the contacts with adjacent molecules of the same nascent groove are formed); at a "fluted" surface, all the molecules of the new layer interact weakly between themselves.

In the first case, the energy barrier retarding the ice initiation is small – it is determined only by the landing of the first molecule in the groove, and the free energy of this barrier is $G^{\#}_{groove} = 2B_1 + \Delta\mu \approx 2B_1$ (in essence, this is an initiation barrier for a 1D system, which, unlike those in the 2D and 3D systems, does not contain large (when $\Delta\mu \to 0$) terms proportional to $B\left(\frac{B}{-\Delta\mu}\right)$); thus, ice can form upon grooved surfaces even at exactly $0°C$. After the initiation, ice grows along the groove quickly and independently of the growth of ice in other grooves. In the second case, the landing of one molecule does not depend on the landing of another, so that the initiation is not at all retarded by the energy barriers, and the ice grows randomly and quickly.

In other words, the loss of surface energy on the perimeter of the growing ice layer is reduced by the corrugated surface and completely eliminated by the fluted one (compare Fig. 3B and 3C with Fig.

3A), and just this drastic reduction of the energy loss is able to equalize the freezing temperature with 0°C.

As to the ice growth at bent surfaces, the freezing at a once-bent surface (Fig. 3D) is similar to that at the "corrugated" one, while the freezing at a twice-bent surface (Fig. 3E) is similar to that at the "fluted" surface.

Unfortunately, I did not manage to find a consideration of any of these interesting cases of crystal initiation in the literature, but it is rather simple to estimate the rates of these processes at "corrugated" and "fluted" surfaces. Since new molecules freeze to different places of a sufficiently large ice surface almost independently, with very small or no barriers against the initiation, the time of freezing of one new layer of waters to the ice is practically the same as the time of freezing of one molecule, estimated by Eq. (5). So, in 1 second, ~100000· $(\frac{\Delta T}{1^\circ})$ new layers of $H_2O$ molecules join the ice, and in a minute the ice adds a layer of ~$(\frac{\Delta T}{1^\circ})$ mm thick.

The above mentioned "corrugated" or "fluted" walls may possibly be present in various objects, including the dust and living cells and blood vessels – although for now, I can only guess how and by what they are formed.

Quite possible that just blocking such ice-initiating surfaces prevents the near-wall ice formation in living cells [29], where the so-called antifreeze proteins (AFPs) [30] (also called ice-binding proteins (IBPs) [31] since they bind to ice [32]) somehow struggle against the ice formation upon overcooling.

To conclude this part, it makes sense to mention the following.

Above, I have noted that the initiation of ice formation without supercooling the water to at least -35°C (see Table 1) is possible only in the presence of ice-binding surfaces.

On the contrary, to initiate ice melting, no overheating is necessary and no "water-binding surfaces" are required since the air (or rarefied water vapor) is "water-binding" by itself (Fig. 5A). This also manifests itself in the fact that ice has a surface similar to supercooled water [33], while neither air nor water vapor is "ice-binding" (Fig. 5B).

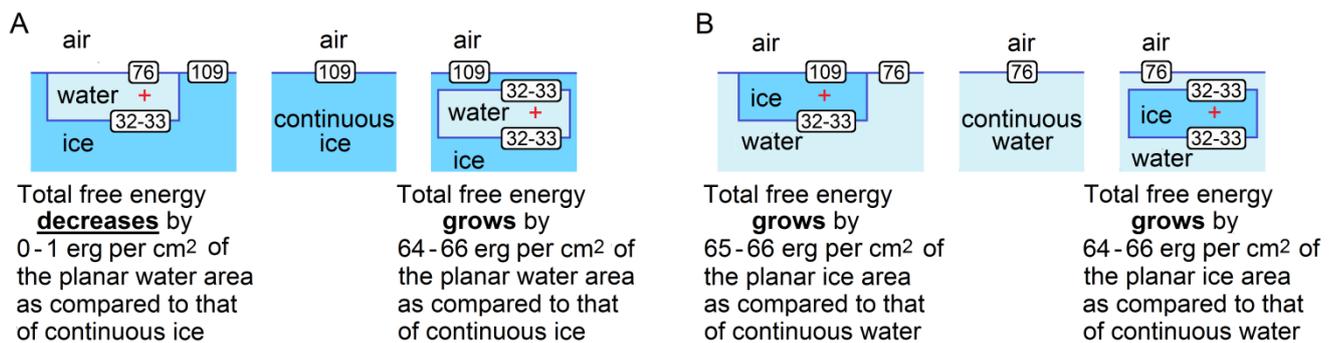

FIG. 5. Schematic presentation of surface free energies (in erg/cm$^2$) for a water puddle on the ice surface and water within ice (A) and for a piece of ice on the water surface and within water (B). The surface free energies are taken from [9, 10, 34, 35].

## Conclusions

The analysis carried out in this work shows that water cannot freeze at zero temperature if it does not simultaneously evaporate from the surface (which can take place in an open vessel and at the air humidity not too close to 100%), or (when the vessel or container is closed – as, for example, a living cell or a blood vessel) if the container does not have specially "grooved" or "bent" (Figs. 3B - 3D) ice-binding surfaces.

## Acknowledgements

I am grateful to M.D. Frank-Kamenetsky, E.G. Malenkov, E.I. Shahnovich, G. Fermi, D. Fraenkel for discussion of the problem, to A.Yu. Grosberg for valuable advices, to A.N. Gavrishev for discussion of meteorological data, to B.S. Melnik for discussing of biophysical data, and to E.V.

Serebrova for editing the manuscript. I am grateful to the RAS Program on Fundamental Research (grant № 01201358029) and the Russian Foundation for Basic Research (grant No. 19–04–00420) for financial support.